# Nature of Electric and Magnetic Dipoles Gleaned from the Poynting Theorem and the Lorentz Force Law of Classical Electrodynamics

Masud Mansuripur

College of Optical Sciences, The University of Arizona, Tucson, Arizona 85721, USA

**Abstract**. Starting with the most general form of Maxwell's macroscopic equations in which the free charge and free current densities, $\rho_{\text{free}}$ and $J_{\text{free}}$, as well as the densities of polarization and magnetization, $P$ and $M$, are arbitrary functions of space and time, we compare and contrast two versions of the Poynting vector, namely, $S=\mu_o^{-1}E\times B$ and $S=E\times H$. Here $E$ is the electric field, $H$ the magnetic field, $B$ the magnetic induction, and $\mu_o$ the permeability of free space. We argue that the identification of one or the other of these Poynting vectors with the rate of flow of electromagnetic energy is intimately tied to the nature of magnetic dipoles and the way in which these dipoles exchange energy with the electromagnetic field. In addition, the manifest nature of both electric and magnetic dipoles in their interactions with the electromagnetic field has consequences for the Lorentz law of force. If the conventional identification of magnetic dipoles with Amperian current loops is extended beyond Maxwell's macroscopic equations to the domain where energy, force, torque, momentum, and angular momentum are active participants, it will be shown that "hidden energy" and "hidden momentum" become inescapable consequences of such identification with Amperian current loops. Hidden energy and hidden momentum can be avoided, however, if we adopt $S=E\times H$ as the true Poynting vector, and also accept a generalized version of the Lorentz force law. We conclude that the identification of magnetic dipoles with Amperian current loops, while certainly acceptable within the confines of Maxwell's macroscopic equations, is inadequate and leads to complications when considering energy, force, torque, momentum, and angular momentum in electromagnetic systems that involve the interaction of fields and matter.

**1. Introduction**. The conventional view of Maxwell's *macroscopic* equations is that they represent nothing new, interesting, or puzzling beyond the fundamental laws of electrodynamics, which laws are elegantly codified in Maxwell's *microscopic* equations plus the Lorentz law of force. The standard argument goes as follows: Substitute for $-\nabla\cdot P$ (i.e., the negative divergence of polarization) the density of bound electric charges, then treat $\partial P/\partial t$ and $\mu_o^{-1}\nabla\times M$ as the densities of bound electric current associated with polarization and magnetization, respectively, and you will arrive at Maxwell's microscopic equations, which relate the electromagnetic (EM) fields, $E(r,t)$ and $B(r,t)$, to their sources, namely, the charge- and current-density distributions [1,2].

The above would be a perfectly sensible viewpoint if all that one cared about were the relating of the fields to their sources, which, after all, is the essence of Maxwell's equations. But there is more to classical electrodynamics than Maxwell's equations per se. For example, there is the matter of relativistic invariance, which demands that the bound electric charge density, $\rho_{\text{e\_bound}}=-\nabla\cdot P$, together with the bound electric current density, $J_{\text{e\_bound}}=\partial P/\partial t + \mu_o^{-1}\nabla\times M$, form a 4-vector. This requirement imposes the well-known restriction on $P(r,t)$ and $M(r,t)$ that, in the Minkowski space-time, they must jointly form a 2$^{\text{nd}}$ rank tensor, which transforms between inertial frames in accordance with the Lorentz transformation rules [3,4].

Then there is the matter of EM energy, which Maxwell's equations do not address directly, and whose basic characteristics are usually inferred from arguments involving force, torque, mechanical work, energy conservation, and other "reasonable" assumptions. In fact, one set of such reasonable assumptions leads to the definition of the Poynting vector as $S_1=\mu_o^{-1}E\times B$ [2], while another set leads to $S_2=E\times H$ [1,5]. If one insists on reducing $P$ and $M$ to $\rho_{\text{e\_bound}}$ and $J_{\text{e\_bound}}$, then one will find the Poynting vector to be none other than $S_1$; see Sec. 3. However, as the arguments of Secs. 4 and 5 demonstrate, adoption of $S_1$ as the Poynting vector would lead to the appearance of a certain amount of "hidden" energy in magnetic media [6,7], which is, at best, inconvenient baggage to drag along, and, at worst, incorrect physics.

There is also the matter of the Lorentz law of force, which prescribes the force exerted by the local $E$ and $B$ fields on charge- and current-density distributions. Again, if one insists on reducing $P$ and $M$ to $\rho_{\text{e\_bound}}$ and $J_{\text{e\_bound}}$, one will find that a certain amount of EM momentum routinely disappears from the electromagnetic field in the presence of magnetic materials [7-10]. The choice is then between accepting the notion of hidden momentum [8,9,11], or circumventing the problem altogether by modifying the Lorentz force law in the presence of magnetic media [12-17]. We will address these choices in Sections 6 and 7.

The question that emerges from these discussions is whether or not magnetic dipoles should be treated as Amperian current loops under all circumstances. The nuanced answer offered in the present paper maintains that, within the confines of Maxwell's macroscopic equations, one can, most definitely, treat magnetic dipoles as Amperian current loops. [This is not a new revelation, but simply what Maxwell's equations have taught us all along with regard to treating polarization $P(r,t)$ and magnetization $M(r,t)$: introduce bound charge and bound current densities into the equations, and you will not have to deal with $P$ and $M$ again.] However, when a magnetic dipole attempts to exchange energy with the EM field, its putative equivalence to an ordinary current loop reveals the limitations of the concept. Can a magnetic dipole take up energy from the local $E$-field without showing any outward signs of this exchange (e.g., a modified dipole moment)? Or, alternatively, can the dipole lend energy to the EM field as necessary, perhaps for an indefinite length of time, without breaking down? These are questions that we will address in Sec. 8, in a critical examination of the Amperian current loop model of magnetic dipoles.

If the Amperian current loop is not the proper model for discussing the exchange of energy and momentum between the fields and the dipoles, then what would be the vehicle of choice for addressing such questions? One answer is that, much in the same vein as Maxwell's equations themselves, perhaps such models are ultimately unnecessary [2]. All that is needed for handling the most general problems involving energy, momentum, force, and torque in classical electrodynamics is a complete and consistent set of rules (i.e., postulates) that would describe such exchanges between the fields and the dipoles.

As we have shown elsewhere [18,19], one such approach postulates the Poynting vector as $S(r,t) = E(r,t) \times H(r,t)$, the electromagnetic linear and angular momentum densities as $p_{EM}(r,t) = S(r,t)/c^2$ and $L_{EM}(r,t) = r \times p_{EM}(r,t)$, and the generalized Lorentz laws of force and torque as given by Eqs. (12a) and (12b) below. These postulates, in conjunction with Maxwell's macroscopic equations, form a complete and consistent basis for analyzing all of classical electrodynamics in compliance with the conservation principles as well as with the special theory of relativity; never again will one need to resort to such artificial devices as hidden energy [7], hidden momentum [8,9], pseudo-momentum [20,21], or special surface forces [21,22], in order to conserve energy, momentum, and angular momentum.

To compare and contrast the approach outlined in the preceding paragraph with the traditional one that insists on reducing $P$ and $M$ to bound charge and bound current densities, is the underlying objective of the present paper. In Sec. 9 we will show the extent to which the Amperian current loop model retains its value as a pedagogical tool; as a physical model for magnetic dipoles, however, we believe the ordinary current loop is deficient and needs to be replaced with the aforementioned postulates. The quantum nature of magnetic dipoles – rooted in the spin and orbital angular momenta of matter's constituents – is ultimately responsible, of course, for their observed behavior under external EM fields. We will not address the quantum-mechanical issues in this paper, however, as our goal is to critically examine the consistency of the classical electrodynamic theory of ponderable media.

It must be emphasized that our intent here is *not* to somehow distinguish a "quantum-mechanical" magnetic dipole from a classical "Amperian current loop" dipole. In the real world, there is only one magnetic dipole, whose associated density appears as $M(r,t)$ in Maxwell's macroscopic equations. No matter how these real-world dipoles are constructed, they exchange energy and momentum with the EM field in a certain way; that reality must be represented by Maxwell's equations, of course, and also by the auxiliary equations that describe the time-rate-of-exchange of EM energy with $M(r,t)$, as well as the force of the EM field experienced by $M(r,t)$. The fact that Maxwell's equations are classical, whereas the spin and orbital magnetic moments are quantum-mechanical in nature, does not have to enter the discussion at all; Maxwell's equations must, at some level, reflect the real world, irrespective of how that world happens to have been constructed.

Also evident in the generalized Lorentz force and torque expressions, Eqs. (12a) and (12b), is a certain departure of the electric dipoles from their expected behavior based on the traditional model of bound charge and bound current densities. It will be pointed out in Sec. 10 that the response of the electric dipoles of a continuous medium to the local $E$-field gives rise to force and torque distributions throughout the material medium that differ substantially from the corresponding distributions expected from bound charges. Once again, the essential quantum-mechanical nature of atomic/molecular dipoles is the likely cause of this departure from the conventional bound-charge model. The difference between the two formulations of the Lorentz force felt by electric dipoles has received scant attention in the literature, presumably because the total (i.e., integrated) force as well as the total torque on a given object turn out to be identical in the two formulations. Nevertheless, the predicted *distributions* of EM force and torque throughout the volume of a soft object should be amenable to experimental verification; this would be a necessary step in the direction of establishing the validity of the generalized Lorentz law of Eqs. (12a) and (12b).

**2. Maxwell's macroscopic equations and the constitutive relations**. In the MKSA system of units, Maxwell's macroscopic equations are written [1-7]:

$$\boldsymbol{\nabla}\cdot\boldsymbol{D}=\rho_{\text{free}}, \tag{1a}$$

$$\boldsymbol{\nabla}\times\boldsymbol{H}=\boldsymbol{J}_{\text{free}}+\partial\boldsymbol{D}/\partial t, \tag{1b}$$

$$\boldsymbol{\nabla}\times\boldsymbol{E}=-\partial\boldsymbol{B}/\partial t, \tag{1c}$$

$$\boldsymbol{\nabla}\cdot\boldsymbol{B}=0. \tag{1d}$$

In these equations, the electric displacement $\boldsymbol{D}$ and the magnetic induction $\boldsymbol{B}$ are related to the polarization $\boldsymbol{P}$ and magnetization $\boldsymbol{M}$ via the identities

$$\boldsymbol{D}=\varepsilon_{\text{o}}\boldsymbol{E}+\boldsymbol{P}, \tag{2a}$$

$$\boldsymbol{B}=\mu_{\text{o}}\boldsymbol{H}+\boldsymbol{M}. \tag{2b}$$

In the general discussions that follow, we will not restrict our attention to any specific constitutive relations for $\boldsymbol{P}$ and $\boldsymbol{M}$, keeping them as arbitrary functions of space and time. Thus, for example, $\boldsymbol{P}$ and $\boldsymbol{M}$ could be dependent on local EM fields, or they could react to fields in distant locations; they could be linear or nonlinear functions of these fields, or they may not depend on the fields at all, being driven by other mechanisms (e.g., mechanical, chemical). Moreover, $\boldsymbol{P}$ and $\boldsymbol{M}$ may exhibit dispersion (spatial as well as temporal), loss, birefringence, optical activity, hysteresis, and so on. Bear in mind that Maxwell's equations do not tell $\boldsymbol{P}(\boldsymbol{r},t)$ and $\boldsymbol{M}(\boldsymbol{r},t)$ how to vary in space and time; the behavior of $\boldsymbol{P}$ and $\boldsymbol{M}$ is determined either independently of the $\boldsymbol{E}$ and $\boldsymbol{B}$ fields, or through some set of constitutive relations, neither of which is under the control of Maxwell's equations.

One can *define* bound electric charge and current densities, $\rho_{\text{e\_bound}}=-\boldsymbol{\nabla}\cdot\boldsymbol{P}(\boldsymbol{r},t)$ and $\boldsymbol{J}_{\text{e\_bound}}=\partial\boldsymbol{P}(\boldsymbol{r},t)/\partial t$, arising from the polarization $\boldsymbol{P}$, as well as a bound electric current density $\boldsymbol{J}_{\text{e\_bound}}=\mu_{\text{o}}^{-1}\boldsymbol{\nabla}\times\boldsymbol{M}$, that gives rise to the magnetization $\boldsymbol{M}$ [12]. Subsequently, the macroscopic equations (1) may be written in the following equivalent way:

$$\varepsilon_{\text{o}}\boldsymbol{\nabla}\cdot\boldsymbol{E}=\rho_{\text{free}}+\rho_{\text{e\_bound}}, \tag{3a}$$

$$\boldsymbol{\nabla}\times\boldsymbol{B}=\mu_{\text{o}}(\boldsymbol{J}_{\text{free}}+\boldsymbol{J}_{\text{e\_bound}})+\mu_{\text{o}}\varepsilon_{\text{o}}\partial\boldsymbol{E}/\partial t, \tag{3b}$$

$$\boldsymbol{\nabla}\times\boldsymbol{E}=-\partial\boldsymbol{B}/\partial t, \tag{3c}$$

$$\boldsymbol{\nabla}\cdot\boldsymbol{B}=0. \tag{3d}$$

The above are the standard *microscopic* equations of Maxwell, which relate the $\boldsymbol{E}$ and $\boldsymbol{B}$ fields to the total electric charge and current density distributions, namely, $\rho_{\text{total}}(\boldsymbol{r},t)=\rho_{\text{free}}+\rho_{\text{e\_bound}}=\rho_{\text{free}}(\boldsymbol{r},t)-\boldsymbol{\nabla}\cdot\boldsymbol{P}(\boldsymbol{r},t)$, and $\boldsymbol{J}_{\text{total}}=\boldsymbol{J}_{\text{free}}+\boldsymbol{J}_{\text{e\_bound}}=\boldsymbol{J}_{\text{free}}+\partial\boldsymbol{P}/\partial t+\mu_{\text{o}}^{-1}\boldsymbol{\nabla}\times\boldsymbol{M}$. Note that, in arriving at Eqs. (3), no assumptions have been made about the nature of $\boldsymbol{P}(\boldsymbol{r},t)$ and $\boldsymbol{M}(\boldsymbol{r},t)$, beyond the fact that these are piecewise continuous and differentiable functions of the space and time coordinates, $\boldsymbol{r}$ and $t$.

It is perhaps best to imagine an "army of ants" controlling the spatio-temporal distributions $\boldsymbol{P}(\boldsymbol{r},t)$ and $\boldsymbol{M}(\boldsymbol{r},t)$. The ants could be programmed to move the dipoles around, to increase or decrease the strengths of the dipoles at each point in space as functions of time, and/or to rotate the dipoles with or without simultaneously modifying their strengths. The ants could be programmed to make their actions contingent upon the local values of the $\boldsymbol{E}$ and $\boldsymbol{B}$ fields; they could respond to the field values at remote locations; or they could ignore the fields altogether. Such is the power of Maxwell's macroscopic equations, that, within a given inertial frame, they impose no restrictions whatsoever on the way that $\boldsymbol{P}$ and $\boldsymbol{M}$ are distributed throughout space and time.

In a ferromagnetic (or ferroelectric) material, for instance, the dipoles are in contact with their neighboring dipoles; quantum-mechanical exchange-coupling makes the dynamic behavior of one dipole dependent on the state of all the other dipoles in the system. Each dipole in this case is driven by a combination of the local $E$- and/or $H$-field, by the direct force of its nearest neighbors, and by the indirect force of other, far-away dipoles. If there is hysteresis in the system, the dynamics of each dipole will also depend on the history of magnetization (or polarization) everywhere in the material medium. In such situations, the local fields do not govern the dynamics of individual dipoles, but "ants," programmed to collect information about the past and present states of the entire system, will be able to drive the dipoles in agreement with the real-world behavior of the material under the action of external fields.

**3. Alternative expressions for the Poynting vector**. Starting with Eqs. (3), if we dot-multiply $\boldsymbol{E}(\boldsymbol{r},t)$ into Eq. (3b), and $\boldsymbol{B}(\boldsymbol{r},t)$ into Eq. (3c), and proceed to subtract the latter from the former, we find the following relations involving the "Poynting vector" $\boldsymbol{S}_1(\boldsymbol{r},t)=\mu_{\text{o}}^{-1}\boldsymbol{E}(\boldsymbol{r},t)\times\boldsymbol{B}(\boldsymbol{r},t)$ and the time-rate-of-change of the field energy density $\mathcal{E}_{\text{1-field}}(\boldsymbol{r},t)$:

$$\boldsymbol{\nabla}\cdot\boldsymbol{S}_1(\boldsymbol{r},t)=-\partial\mathcal{E}_{\text{1-field}}(\boldsymbol{r},t)/\partial t, \tag{4a}$$

$$\frac{\partial\mathcal{E}_{\text{1-field}}(\boldsymbol{r},t)}{\partial t}=\boldsymbol{E}\cdot\boldsymbol{J}_{\text{total}}+\partial(\tfrac{1}{2}\varepsilon_\text{o}E^2+\tfrac{1}{2}\mu_\text{o}^{-1}B^2)/\partial t. \tag{4b}$$

Alternatively, if we do *not* invoke bound currents, and simply perform the corresponding operations on the original macroscopic equations, Eqs. (1b) and (1c), we find the Poynting vector to be $\boldsymbol{S}_2(\boldsymbol{r},t)=\boldsymbol{E}(\boldsymbol{r},t)\times\boldsymbol{H}(\boldsymbol{r},t)$, while the time-rate-of-change of the field energy density becomes

$$\frac{\partial\mathcal{E}_{\text{2-field}}(\boldsymbol{r},t)}{\partial t}=\boldsymbol{E}\cdot\boldsymbol{J}_{\text{free}}+\boldsymbol{E}\cdot\partial\boldsymbol{D}/\partial t+\boldsymbol{H}\cdot\partial\boldsymbol{B}/\partial t=\boldsymbol{E}\cdot\boldsymbol{J}_{\text{free}}+\boldsymbol{E}\cdot\partial\boldsymbol{P}/\partial t+\boldsymbol{H}\cdot\partial\boldsymbol{M}/\partial t+\partial(\tfrac{1}{2}\varepsilon_\text{o}E^2+\tfrac{1}{2}\mu_\text{o}H^2)/\partial t. \tag{5}$$

Interpretation of the various terms on the right-hand-side of Eq. (5), although straightforward, requires some attention to detail, as the equation does not depend on any specific model of the material medium. The term $\boldsymbol{E}\cdot\boldsymbol{J}_{\text{free}}$, for example, is the time-rate-of-flow of EM energy into or out of the carriers of free current, depending on whether the sign of the dot-product is positive or negative. The equation does not tell what happens to the energy when it disappears into the charge carriers (e.g., producing heat, accelerating the carriers, enabling chemical reactions, etc.), nor does it specify where the energy comes from when the carriers deliver the energy to the EM field.

Similarly, the second term on the right-hand-side of Eq. (5) states that the electric dipoles at a given point in space-time absorb energy from or deliver energy to the field, at the rate of $\boldsymbol{E}\cdot\partial\boldsymbol{P}/\partial t$, irrespective of how the change in $\boldsymbol{P}$ is brought about. It matters not whether it is the magnitude or the direction of $\boldsymbol{P}$ that changes with time; it is immaterial whether $\boldsymbol{P}$ is controlled by the local EM field, by fields in remote locations, or simply by the "ants" mentioned in Sec. 2, with no regard whatsoever for the strength or the direction of the field. The equation does not tell where the energy goes to when taken up by the dipoles, nor where it comes from when given by the dipoles to the field; all one knows is that the "ants" pick up the energy when the dot-product $\boldsymbol{E}\cdot\partial\boldsymbol{P}/\partial t$ is positive, and that, somehow, they find the resources to produce the energy at the desired rate when the dot-product is negative.

The same considerations apply to the third term on the right-hand-side of Eq. (5), $\boldsymbol{H}\cdot\partial\boldsymbol{M}/\partial t$, with the obvious distinction that here $\boldsymbol{H}(\boldsymbol{r},t)$ is the field that acts on the local magnetic dipoles. Here, it is irrelevant how the *E*-field is distributed in space-time, or whether $\partial\boldsymbol{M}/\partial t$ is the result of a change in the magnitude and/or orientation of the magnetic dipoles; all that Eq. (5) reveals is that the "ants" pick up energy from the EM field at the rate of $\boldsymbol{H}\cdot\partial\boldsymbol{M}/\partial t$ when the dot-product is positive, and that they somehow manage to deliver energy to the field at the specified rate when the dot-product is negative. The ants may store the energy that they pick up at one instant, then return it to the field at later times, or they could burn the absorbed energy on the spot, or they may choose to exchange energy with other ants in the neighborhood. Each such possibility may represent one model of the material medium or another, but none has any relevance to the implications of Eq. (5) with regard to the exchange of energy between the EM field and the material medium.

Some authors have suggested that the terms $\boldsymbol{E}\cdot\partial\boldsymbol{P}/\partial t$ and $\boldsymbol{H}\cdot\partial\boldsymbol{M}/\partial t$ must assume the form of complete differentials before they can be cast as $\partial\mathcal{E}_{\text{field}}/\partial t$. This, they have argued, happens, for example, when $\boldsymbol{P}(\boldsymbol{r},t)$ and $\boldsymbol{M}(\boldsymbol{r},t)$ are proportional to the local fields $\boldsymbol{E}(\boldsymbol{r},t)$ and $\boldsymbol{H}(\boldsymbol{r},t)$, respectively [23,24]. Others have demonstrated the possibility of arriving at the Poynting theorem in the form of Eqs. (4) under the broad – but nonetheless restrictive– condition that, in its response to the *E* and *B* fields, the material medium be spatially and/or temporally dispersive, albeit in a linear way [25]. Standing opposed to these authors, we believe that Eq. (5) is valid for *any* kind of material medium, so long as the exchange of energy between the EM field and elementary electric and magnetic dipoles follows the precepts expounded in the preceding paragraphs.

Thus, based solely on the formal derivation of Eqs. (4) and (5) outlined in the beginning of the present section, if we accept that the Poynting vector is $\boldsymbol{E}(\boldsymbol{r},t)\times\boldsymbol{H}(\boldsymbol{r},t)$ rather than $\mu_\text{o}^{-1}\boldsymbol{E}(\boldsymbol{r},t)\times\boldsymbol{B}(\boldsymbol{r},t)$, we must reject Eq. (4b) in favor of Eq. (5). In particular, we must recognize that the bound current of magnetization, $\boldsymbol{J}_{\text{e\_bound}}=\mu_\text{o}^{-1}\boldsymbol{\nabla}\times\boldsymbol{M}$, behaves differently than the other two types of current, namely, $\boldsymbol{J}_{\text{free}}$ and $\boldsymbol{J}_{\text{e\_bound}}=\partial\boldsymbol{P}(\boldsymbol{r},t)/\partial t$. To appreciate the distinction, note that, in the Amperian loop model, $\mu_\text{o}^{-1}\boldsymbol{\nabla}\times\boldsymbol{M}$ is directly related to the internal currents of the magnetic dipoles, which currents are entirely responsible for the existence of $\boldsymbol{M}$. For example, for a thin solid disk, having a uniform magnetization $\boldsymbol{M}$ perpendicular to the plane of the disk and constant in time, $\mu_\text{o}^{-1}\boldsymbol{\nabla}\times\boldsymbol{M}$ is a constant surface current density circulating around the edge of the disk. Now, if the integral of $\boldsymbol{E}\cdot(\mu_\text{o}^{-1}\boldsymbol{\nabla}\times\boldsymbol{M})$ over the disk's volume happens to be non-zero, energy must be flowing in or out of the disk in accordance with Eq. (4b). In contrast, since the magnetization $\boldsymbol{M}$ of the disk is not changing with time, there can be no exchange of energy according to the $\boldsymbol{H}\cdot\partial\boldsymbol{M}/\partial t$ formulation of Eq. (5); clearly, the two formulations predict very different behaviors. If one believes, erroneously, that a current density $\boldsymbol{J}$, irrespective of its origins, must exchange energy with the local *E*-field at the rate of $\boldsymbol{E}\cdot\boldsymbol{J}$, then the $\boldsymbol{H}\cdot\partial\boldsymbol{M}/\partial t$ formulation cannot possibly be correct. However, since we contend that $H \cdot \partial M / \partial t$ is in fact the correct energy-exchange rate, we are left with no choice but to abandon the belief in the $E \cdot J$ formulation where $J_{\text{e\_bound}} = \mu_o^{-1} \nabla \times M$ is concerned.

Here we have a strong statement concerning the nature of magnetization $M(r,t)$. The quantum nature of spin and orbital magnetic dipole moments provides no *a priori* grounds for believing that magnetic dipoles should in every respect behave like Amperian current loops. Whereas the behavior of $M(r,t)$ in the context of Maxwell's equations alone *can* be represented by an equivalent bound current density $J_{\text{e\_bound}} = \mu_o^{-1} \nabla \times M$, when it comes to exchanging energy with the EM field, $M(r,t)$ turns out to behave differently. These are sufficient grounds for stating that the form of the Poynting vector provides a strong clue as to the nature of the magnetic dipoles. If the Poynting vector happens to be $E(r,t) \times H(r,t)$, then the Amperian current loops commonly associated with $M(r,t)$ will *not* be ordinary current loops at all; these will have to have their own quantum-mechanical properties and, in particular, their own way of exchanging energy with the EM field.

**4. The case for $S = E \times H$.** With reference to Fig. 1, consider a slab of homogeneous, isotropic, transparent, dispersionless, magnetic dielectric, which has $\varepsilon = \mu > 1$. Here $\varepsilon$ and $\mu$ are the relative permittivity and permeability of the slab, respectively. A light pulse arriving at normal incidence on the entrance facet will enter the slab with no reflection losses whatsoever. In other words, the slab is perfectly impedance-matched to the free-space. Now, Maxwell's boundary conditions require the continuity of tangential $E$ and $H$ fields, which makes the incident energy flux in the free space precisely equal to the transmitted energy flux within the slab, if the Poynting vector happens to be $S = E \times H$ [1,5,12]. In contrast, if $S$ were taken to be $\mu_o^{-1} E \times B$ [2], we will see in Sec. 5 that some energy would have to be "borrowed" from the magnetic dipoles residing at the entrance facet. This borrowed energy will, of course, return to the magnetic dipoles at the exit facet, once the light finally leaves the slab.

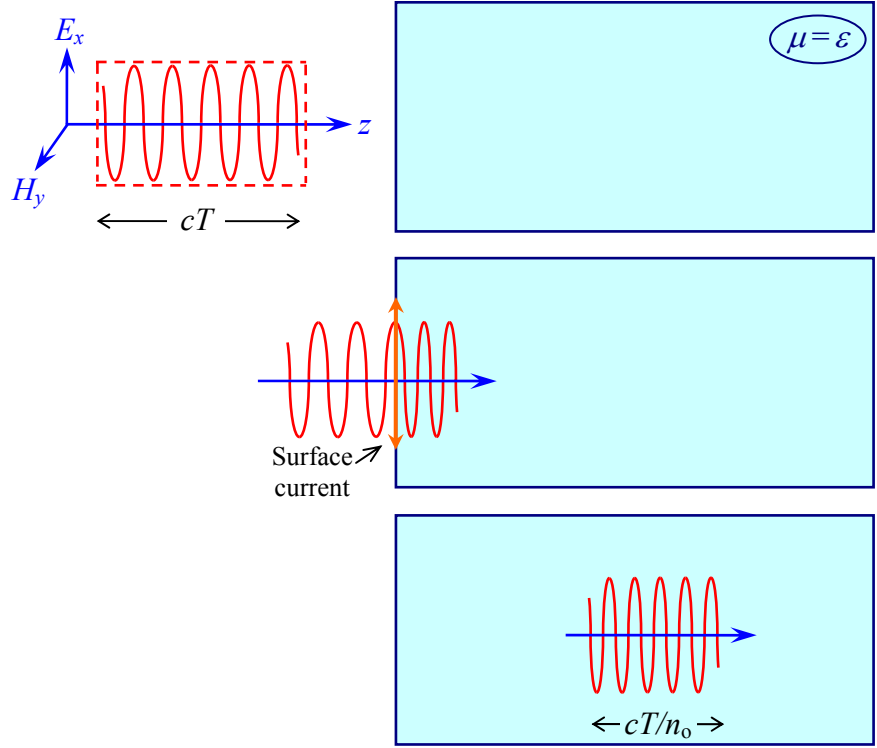

**Fig. 1**. A rectangular pulse of light enters a transparent, dispersionless, homogeneous, isotropic, linear, magnetic dielectric whose dielectric permittivity $\varepsilon$ and magnetic permeability $\mu$ happen to be equal to each other. From top to bottom: before, during, and after the pulse entry into the slab. The sharp discontinuity of $M(r,t) = \mu_o(\mu-1) H(r,t)$ at the entrance facet produces a bound electric surface current at this facet during the time interval $-T \le t \le 0$ when the pulse enters the slab (middle frame). Elsewhere, the total bound current density is given by $J_{\text{e\_bound}} = \partial P / \partial t + \mu_o^{-1} \nabla \times M$. The leading edge of the pulse exerts a force on the material medium that tends to push the local atoms/molecules forward; the trailing edge exerts an equal force in the opposite direction. The conventional Lorentz force $J \times B$ gives rise to a negative (pull) force on the entrance facet of the slab by the action of the $B$-field of the light pulse on the induced surface current. This surface force turns out to be equal and opposite to the force exerted by the leading edge of the pulse on the interior atoms/molecules of the slab. Consequently, if the conventional Lorentz law happens to be operative, no net mechanical momentum will be transferred from the pulse to the slab.

Defining the rectangular pulse $\Pi(t)$ as a unit-amplitude function of given duration $T$ confined to the interval $0 \leq t \leq T$, a linearly-polarized light pulse having center-frequency $\omega_o$, $E$-field amplitude $E_o\hat{x}$, and propagation direction $\hat{z}$, will have the following $E$- and $H$-fields within a medium of permittivity $\varepsilon_o\varepsilon$ and permeability $\mu_o\mu$:

$$\boldsymbol{E}(\boldsymbol{r},t) = E_o \sin[\omega_o(t - n_o z/c) + \phi_o] \Pi(t - n_o z/c)\hat{x}, \tag{6a}$$

$$\boldsymbol{H}(\boldsymbol{r},t) = \sqrt{\varepsilon_o\varepsilon/\mu_o\mu}\, E_o \sin[\omega_o(t - n_o z/c) + \phi_o] \Pi(t - n_o z/c)\hat{y}. \tag{6b}$$

Here $n_o = \sqrt{\mu\varepsilon}$ is the refractive index of the material medium, $c = 1/\sqrt{\mu_o\varepsilon_o}$ is the speed of light in vacuum, and $Z_o = \sqrt{\mu_o/\varepsilon_o}$ is the impedance of the free space. The pulse duration $T$ and the phase angle $\phi_o$ are arbitrary constants. The following expressions for the Poynting vector, its divergence, and the time-rate-of-change of the field's energy density are valid when the pulse is either entirely in the free space, where $\varepsilon = \mu = 1$, or entirely within the slab.

$$\boldsymbol{S}(\boldsymbol{r},t) = \boldsymbol{E}(\boldsymbol{r},t) \times \boldsymbol{H}(\boldsymbol{r},t) = \sqrt{\varepsilon_o\varepsilon/\mu_o\mu}\, E_o^2 \sin^2[\omega_o(t - n_o z/c) + \phi_o]\Pi(t - n_o z/c)\hat{z}; \tag{7a}$$

$$\boldsymbol{\nabla}\cdot\boldsymbol{S}(\boldsymbol{r},t) = \partial S_z(\boldsymbol{r},t)/\partial z = -\varepsilon_o\varepsilon E_o^2 \{\omega_o \sin[2\omega_o(t - n_o z/c) + 2\phi_o]\Pi(t - n_o z/c)$$
$$+ \sin^2[\omega_o(t - n_o z/c) + \phi_o][\delta(t - n_o z/c) - \delta(t - T - n_o z/c)]\}; \tag{7b}$$

$$\partial\mathcal{E}(\boldsymbol{r},t)/\partial t = \boldsymbol{E}\cdot\partial\boldsymbol{D}/\partial t + \boldsymbol{H}\cdot\partial\boldsymbol{B}/\partial t = \varepsilon_o\varepsilon E_o^2\, \partial\{\sin^2[\omega_o(t - n_o z/c) + \phi_o]\Pi(t - n_o z/c)\}/\partial t$$
$$= \varepsilon_o\varepsilon E_o^2\{\omega_o \sin[2\omega_o(t - n_o z/c) + 2\phi_o]\Pi(t - n_o z/c) + \sin^2[\omega_o(t - n_o z/c) + \phi_o][\delta(t - n_o z/c) - \delta(t - T - n_o z/c)]\}. \tag{7c}$$

The two $\delta$-functions appearing in Eqs. (7b) and (7c) correspond to the leading and trailing edges of the rectangular light pulse. Clearly, $\boldsymbol{\nabla}\cdot\boldsymbol{S}(\boldsymbol{r},t) + \partial\mathcal{E}(\boldsymbol{r},t)/\partial t = 0$, confirming energy conservation everywhere at all times.

When the pulse is in the process of entering the slab at its front facet (located at $z = 0$), the $E$-fields on the two sides of the interface ($z = 0^\pm$) are equal, and so are the $H$-fields. Consequently, the Poynting vector $\boldsymbol{S}$ remains continuous at the front facet – and no energy is exchanged between the light pulse and this facet – as the pulse moves smoothly from the vacuum into the slab. While entering the slab ($-T \leq t \leq 0$), the pulse in the vacuum side of the interface, where $n_o = 1$, will have a trailing edge only, corresponding to the term $\delta(t - T - z/c)$ in Eqs. (7b) and (7c), while the pulse inside the slab will have a leading-edge only, corresponding to the term $\delta(t - n_o z/c)$.

**5. The case against $\boldsymbol{S} = \mu_o^{-1}\boldsymbol{E}\times\boldsymbol{B}$.** In this case the Poynting vector, its divergence, and the time-rate-of-change of the field's energy density are given by

$$\boldsymbol{S}(\boldsymbol{r},t) = \mu_o^{-1}\boldsymbol{E}(\boldsymbol{r},t)\times\boldsymbol{B}(\boldsymbol{r},t) = (n_o/Z_o)E_o^2 \sin^2[\omega_o(t - n_o z/c) + \phi_o]\Pi(t - n_o z/c)\hat{z}; \tag{8a}$$

$$\boldsymbol{\nabla}\cdot\boldsymbol{S}(\boldsymbol{r},t) = -\varepsilon_o n_o^2 E_o^2\{\omega_o \sin[2\omega_o(t - n_o z/c) + 2\phi_o]\Pi(t - n_o z/c) + \sin^2[\omega_o(t - n_o z/c) + \phi_o][\delta(t - n_o z/c) - \delta(t - T - n_o z/c)]\}; \tag{8b}$$

$$\partial\mathcal{E}(\boldsymbol{r},t)/\partial t = \boldsymbol{E}\cdot(\partial\boldsymbol{P}/\partial t + \mu_o^{-1}\boldsymbol{\nabla}\times\boldsymbol{M}) + \partial(\tfrac{1}{2}\varepsilon_o E^2 + \tfrac{1}{2}\mu_o^{-1}B^2)/\partial t$$
$$= \tfrac{1}{2}\varepsilon_o(\varepsilon - 1)\partial E^2/\partial t + \tfrac{1}{2}\varepsilon_o\varepsilon(\mu - 1)\partial E^2/\partial t + \tfrac{1}{2}\varepsilon_o\partial E^2/\partial t + \tfrac{1}{2}\mu_o\mu^2\partial H^2/\partial t$$
$$= \varepsilon_o\varepsilon\mu\, \partial E^2(\boldsymbol{r},t)/\partial t. \tag{8c}$$

So long as the pulse remains entirely within one medium or the other, the situation will be similar to that in the preceding section, with $\boldsymbol{\nabla}\cdot\boldsymbol{S}(\boldsymbol{r},t) + \partial\mathcal{E}(\boldsymbol{r},t)/\partial t = 0$ confirming conservation of energy. The difference between the present case and the previous one, however, shows up during the interval $-T \leq t \leq 0$, when the pulse is in the process of entering the slab. As before, the tangential $\boldsymbol{E}$ and $\boldsymbol{H}$ are continuous at the interface, which means that the tangential $\boldsymbol{B}$ has a discontinuity. The Poynting vector, therefore, is not the same on the two sides of the interface, causing the time-rate of arrival of energy from the free-space onto the slab surface to differ from the time-rate of flow of energy into the slab. The gap between these two rates is bridged by the rate at which energy is "extracted" or "borrowed" from the magnetic dipoles located at the entrance facet. Consider the bound surface current at the entrance facet during the period of pulse entry into the material medium, namely,

$$\boldsymbol{J}_{e\_bound} = \mu_o^{-1}\boldsymbol{\nabla}\times\boldsymbol{M} = -(\mu - 1)\sqrt{\varepsilon_o\varepsilon/\mu_o\mu}\, E_o \sin(\omega_o t + \phi_o)\delta(z)\hat{x} = -(n_o - \sqrt{\varepsilon/\mu})(E_o/Z_o)\sin(\omega_o t + \phi_o)\delta(z)\hat{x}. \tag{9}$$

Multiplying this surface current density with the $E$-field at the entrance facet, $E_o \sin(\omega_o t + \phi_o)\hat{x}$, then integrating over $z$, we find the time-rate of energy generation at the entrance facet by the bound surface current of $\boldsymbol{M}(\boldsymbol{r},t)$ to be $(n_o - \sqrt{\varepsilon/\mu})(E_o^2/Z_o)\sin^2(\omega_o t + \phi_o)$. This, of course, is precisely the additional energy flux that is needed to account for the difference, namely, $\mu_o^{-1} \boldsymbol{E} \times \boldsymbol{M}$, between the arrival rate of energy from the vacuum onto the slab's front facet, and the rate at which energy pours into the slab. The magnetic dipoles located at the front facet thus "lend" energy to the pulse as it enters. Carrying this extra energy into the slab, the pulse takes it all the way to the exit facet, at which point it returns the extra energy to the magnetic dipoles of the exit facet, then emerges into the vacuum with the same energy that it had prior to entering the slab.

Note that the bound surface current associated with the curl of $\boldsymbol{M}(\boldsymbol{r},t)$ at the entrance and exit facets of the slab is always present, irrespective of how the Poynting vector is defined. The difference is that, in the $\boldsymbol{S} = \boldsymbol{E} \times \boldsymbol{H}$ formulation, this surface current is *not* permitted to exchange energy with the $E$-field of the light beam; see Eq. (5). In contrast, in the $\boldsymbol{S} = \mu_o^{-1} \boldsymbol{E} \times \boldsymbol{B}$ formulation, the surface current is required to take part in such exchanges following the dictates of Eq. (4b).

**6. Force, torque, and momentum: the conventional approach**. The force exerted by the electromagnetic $\boldsymbol{E}$ and $\boldsymbol{B}$ fields on material media may be obtained from the Lorentz law of force, $\boldsymbol{f} = q(\boldsymbol{E} + \boldsymbol{V} \times \boldsymbol{B})$ [1,2,5]. To convert this expression of force on a point-particle of charge $q$ and velocity $\boldsymbol{V}$, into one of force density $\boldsymbol{F}(\boldsymbol{r},t)$ experienced by a continuum of free and bound charges and currents, one replaces $q$ with the total charge density $\rho_{\text{free}}(\boldsymbol{r},t) - \nabla \cdot \boldsymbol{P}(\boldsymbol{r},t)$, and $q\boldsymbol{V}$ with the total current density $\boldsymbol{J}_{\text{free}}(\boldsymbol{r},t) + \partial \boldsymbol{P}/\partial t + \mu_o^{-1} \nabla \times \boldsymbol{M}$. [In this formulation, the torque density is $\boldsymbol{T}(\boldsymbol{r},t) = \boldsymbol{r} \times \boldsymbol{F}(\boldsymbol{r},t)$; torque density will not enter the following discussions, but is included here for completeness.] In the problem of pulse propagation through the slab depicted in Fig. 1, $\rho_{\text{free}}$ and $\boldsymbol{J}_{\text{free}}$ are absent, and $\nabla \cdot \boldsymbol{P}(\boldsymbol{r},t) = 0$ everywhere. We have

$$\boldsymbol{F}(\boldsymbol{r},t) = (\partial \boldsymbol{P}/\partial t + \mu_o^{-1} \nabla \times \boldsymbol{M}) \times \boldsymbol{B}(\boldsymbol{r},t) = \varepsilon_o(n_o/c)(n_o^2 - 1)E_o^2 \Pi(t - n_o z/c) \{\tfrac{1}{2}\omega_o \sin[2\omega_o(t - n_o z/c) + 2\phi_o]$$
$$+ [\delta(t - n_o z/c) - \delta(t - T - n_o z/c)]\sin^2[\omega_o(t - n_o z/c) + \phi_o]\}\hat{\boldsymbol{z}}. \tag{10}$$

In the above expression of force density, the term that is proportional to $\sin[2\omega_o(t - n_o z/c) + 2\phi_o]$ acts on the volume of material occupied by the entire pulse at any given instant of time. When integrated over the length of the pulse, this term averages out to zero. The remaining two terms, each containing a $\delta$-function, are force densities at the leading and trailing edges of the pulse. To find the total force acting on each edge (per unit cross-sectional area), the $\delta$-functions must be integrated over $z$. Since $\Pi(t - n_o z/c)$ is multiplied into $\delta$-functions whose locations along the $z$-axis coincide with the sharp edges of $\Pi(\cdot)$, the area under each $\delta$-function will be cut in half. Multiplication by $\sin^2[\omega_o(t - n_o z/c) + \phi_o]$, however, has a somewhat subtler effect on the integrals. At first sight, it appears that the leading edge $\delta$-function is multiplied by $\sin^2\phi_o$, that of the trailing edge by $\sin^2(\omega_o T + \phi_o)$. However, any small amount of dispersion in the material would cause the envelope of the pulse to propagate at a different velocity than the sinusoidal carrier, in which case the phase $\phi_o$ will become a rapidly changing function of time. Averaging over all possible values of $\phi_o$ thus produces another factor of $\tfrac{1}{2}$ that multiplies both $\delta$-functions. The net force per unit cross-sectional area at the leading and trailing edges of the pulse is thus found to be $\boldsymbol{f} = \pm \tfrac{1}{4}\varepsilon_o(n_o^2 - 1)E_o^2 \hat{\boldsymbol{z}}$.

In addition to the forces acting on the medium at the leading and trailing edges of the light pulse, there is the force of the $B$-field acting on the surface current density $\boldsymbol{J}_{\text{e\_bound}}$ of Eq. (9) at the front facet of the slab. The discontinuity of $\boldsymbol{B}(\boldsymbol{r},t)$ at the front facet means that the effective $B$-field acting on the surface current must be $\boldsymbol{B}(x,y,z=0,t) = \tfrac{1}{2}\mu_o(\mu+1)H_y(x,y,z=0,t)\hat{\boldsymbol{y}}$. We thus have

$$\boldsymbol{F}_s(\boldsymbol{r},t) = \boldsymbol{J}_{\text{e\_bound}} \times \boldsymbol{B}(x,y,z=0,t) = -\tfrac{1}{2}\varepsilon_o[\varepsilon\mu - (\varepsilon/\mu)]E_o^2 \sin^2(\omega_o t + \phi_o)\delta(z)\hat{\boldsymbol{z}}. \tag{11}$$

When time-averaged and integrated over $z$, the above formula yields a surface force (per unit cross-sectional area) equal to $-\tfrac{1}{4}\varepsilon_o[\varepsilon\mu - (\varepsilon/\mu)]E_o^2 \hat{\boldsymbol{z}}$. For the slab with $\varepsilon = \mu$, this surface force is seen to exactly cancel out the force of the leading edge, resulting in a zero net force experienced by the slab during the entry phase. Of course, once the pulse is fully inside, the forces exerted by the leading and trailing edges of the pulse cancel each other out, with the result that no net mechanical force acts on the slab at any time, either during the entry phase or afterward. Since the Balazs thought experiment [26] requires the electromagnetic momentum inside the medium to be $1/n_o$ times its free-space value, we are thus faced with a case of missing momentum in the amount of $\tfrac{1}{2}\varepsilon_o(1 - n_o^{-1})E_o^2 T \hat{\boldsymbol{z}}$.

**7. Eliminating hidden momentum with the aid of generalized Lorentz force**. Consider the alternative expressions of the EM force and torque densities exerted by the $\boldsymbol{E}$ and $\boldsymbol{H}$ fields on material media [7,12,14]:

$$\boldsymbol{F}(\boldsymbol{r},t) = \rho_{\text{free}}\boldsymbol{E} + \boldsymbol{J}_{\text{free}} \times \mu_o \boldsymbol{H} + (\boldsymbol{P} \cdot \nabla)\boldsymbol{E} + (\partial \boldsymbol{P}/\partial t) \times \mu_o \boldsymbol{H} + (\boldsymbol{M} \cdot \nabla)\boldsymbol{H} - (\partial \boldsymbol{M}/\partial t) \times \varepsilon_o \boldsymbol{E}; \tag{12a}$$

$$\boldsymbol{T}(\boldsymbol{r},t) = \boldsymbol{r} \times \boldsymbol{F}(\boldsymbol{r},t) + \boldsymbol{P}(\boldsymbol{r},t) \times \boldsymbol{E}(\boldsymbol{r},t) + \boldsymbol{M}(\boldsymbol{r},t) \times \boldsymbol{H}(\boldsymbol{r},t). \tag{12b}$$

There are no free charges, nor free currents, in the problem of pulse propagation depicted in Fig. 1; also $(\boldsymbol{P}\cdot\nabla)\boldsymbol{E}$ and $(\boldsymbol{M}\cdot\nabla)\boldsymbol{H}$ vanish everywhere. Furthermore, there are no bound surface charges/currents at the entrance facet of the slab. Therefore, the force density exerted by the light pulse on the slab may be written as follows [16,27-31]:

$$\boldsymbol{F}(\boldsymbol{r},t) = (\partial \boldsymbol{P}/\partial t) \times \mu_0 \boldsymbol{H} - (\partial \boldsymbol{M}/\partial t) \times \varepsilon_0 \boldsymbol{E} = (\varepsilon_0/c)\sqrt{\varepsilon/\mu}\,(\varepsilon+\mu-2)E_0^2\,\Pi(t-n_0z/c)\,\{\tfrac{1}{2}\omega_0\sin[2\omega_0(t-n_0z/c)+2\phi_0]$$
$$+ [\delta(t-n_0z/c) - \delta(t-T-n_0z/c)]\sin^2[\omega_0(t-n_0z/c)+\phi_0]\}\hat{\boldsymbol{z}}. \tag{13}$$

As before, the term that is proportional to $\sin[2\omega_0(t-n_0z/c)+2\phi_0]$ is the volume force density that averages out to zero upon integration over the length of the pulse. The two $\delta$-functions represent the force densities at the leading and trailing edges of the pulse; the total force per unit cross-sectional area acting on each edge is obtained by integrating these $\delta$-functions over $z$. Once again, multiplication by $\Pi(t-n_0z/c)$ cuts the area under each $\delta$-function by half, and multiplication by $\sin^2[\omega_0(t-n_0z/c)+\phi_0]$, when averaged over all possible values of $\phi_0$, produces another factor of ½. The net force per unit cross-sectional area of the leading and trailing edges of the pulse is thus given by

$$\boldsymbol{f} = \pm\tfrac{1}{4}\varepsilon_0[(\varepsilon+\mu-2)/\mu]E_0^2\hat{\boldsymbol{z}}. \tag{14}$$

(This result is in agreement with those obtained by assuming shapes other than rectangular for the light pulse [16].) When the pulse is fully inside the slab, the push of the leading edge on the material medium is precisely balanced by the pull of the trailing edge, resulting in a net total force of zero and, therefore, no change in the mechanical momentum of the slab. However, during the entry phase of the pulse, the leading edge imparts a forward mechanical momentum to the slab that is equal to $\tfrac{1}{4}\varepsilon_0[(\varepsilon+\mu-2)/\mu]E_0^2 T$ per unit cross-sectional area. Since we are presently assuming $\mu=\varepsilon=n_0$, the slab's total acquired mechanical momentum will be $\tfrac{1}{2}\varepsilon_0(1-n_0^{-1})E_0^2 T\hat{\boldsymbol{z}}$. Now, in accordance with the Balazs thought experiment [26,27], the total EM momentum of the light pulse inside the transparent slab must be $\tfrac{1}{2}\varepsilon_0 n_0^{-1}E_0^2 T\hat{\boldsymbol{z}}$. We see that the sum of the mechanical and electromagnetic momenta is given by $\tfrac{1}{2}\varepsilon_0 E_0^2 T\hat{\boldsymbol{z}}$, which is the total momentum of the light pulse before entering the slab. It is thus seen that momentum is conserved and the Balazs requirement is satisfied, without invoking hidden momentum.

**8. Contribution of *E*-field to the internal energy of a magnetic dipole**. Here we try to explain the origin of hidden energy when the local *E*-field is allowed to exchange energy with an Amperian current loop. Let a magnetic point-dipole, $m_0\hat{\boldsymbol{z}}$, be modeled as a tiny current loop sitting in the *xy*-plane at the origin of the coordinate system. The loop's current $I_0$, its surface area $A$, and its dipole moment $m_0$ are related via $m_0=\mu_0 I_0 A$. Assume now that a static *E*-field with components $(E_x, E_y)$ acts on the loop current in such a way that $\oint(\partial\mathcal{E}/\partial t)\,\mathrm{d}\ell = \oint\boldsymbol{E}\cdot\boldsymbol{J}\,\mathrm{d}\ell$ around the loop has a fixed, positive value. This, of course, implies that the *E*-field has a non-zero curl at the origin, which, in turn, requires the existence of a time-varying magnetic field, $\boldsymbol{H}(t)=-H_0 t\hat{\boldsymbol{z}}$ along the z-axis $(\nabla\times\boldsymbol{E}=\mu_0 H_0\hat{\boldsymbol{z}})$. Energy is therefore pouring into this dipole at the fixed rate of $\oint\boldsymbol{E}\cdot\boldsymbol{J}\,\mathrm{d}\ell$, but where does the energy go?

If it were an ordinary current loop, we surely would have expected the energy to go to waste via resistive heating, or, in the absence of resistance, expected it to help accelerate the conduction electrons and, thereby, increase the magnetic dipole moment of the loop. None of these are expected from an actual magnetic dipole. Does a spinning electron really heat up (or cool down) under similar circumstances? Does the electron's magnetic moment rise (or fall) indefinitely, so long as the applied magnetic field keeps its pace of decreasing (or increasing) with time? Such is the nature of hidden energy that first came to light in Sec. 5, and, indeed, it seems implausible that actual magnetic dipoles would behave in such a bizarre way.

However, if the exchange of energy between the field and the magnetic dipole follows the formula $\partial\mathcal{E}/\partial t = \boldsymbol{H}\cdot\partial\boldsymbol{M}/\partial t$, as suggested by Eq. (5), no energy will be given to or taken away from the dipole, so long as its magnetic moment *m* remains unchanged. In this respect, the magnetic dipole would behave similarly to an electric dipole *p*, whose exchange of energy with the external EM field is governed by the formula $\partial\mathcal{E}/\partial t = \boldsymbol{E}\cdot\partial\boldsymbol{P}/\partial t$; see Eq. (5). The only time that the electric (magnetic) dipole gives energy to the field, or takes energy away from the field, is when *p* (*m*) varies with time in the presence of an external field *E* (*H*). Note also that neither dipole, *p* or *m*, is "required" to change in any way in response to the local field. To use once again the metaphor of "ants" mentioned in Sec. 2, only the ants are allowed to move the dipoles around, or to change the magnitude and/or orientation of a given dipole in accordance with their internal programming (i.e., constitutive relations). In other words, whereas the EM field is produced by the dipoles, the dipoles are not "required" to respond to the field. Only when the "ants" choose to increase or decrease the dipole moments, or when they decide to translate or rotate the dipoles, will there be occasion for energy exchange between the EM field and the dipoles. (Contrast this situation with the Amperian current loop model of a magnetic dipole, where the loop could be made to either give or receive energy, even when its dipole moment *m* remains fixed.)

**9. Force exerted by electromagnetic fields on a magnetic dipole**. The physical interpretation of the term $(\boldsymbol{P}\cdot\boldsymbol{\nabla})\boldsymbol{E}$ in Eq. (12a) is straightforward: any differences in the magnitude and/or direction of the *E*-field at the two poles of an electric dipole *p* will result in a net force experienced by the dipole. The same picture does not readily apply to the term $(\boldsymbol{M}\cdot\boldsymbol{\nabla})\boldsymbol{H}$ appearing within the same equation, if one suspects a magnetic dipole to be, at some level, a current loop rather than a pair of closely spaced magnetic monopoles. For simplicity's sake and with reference to Fig. 2(a), let us consider a magnetostatic problem in which the point dipole $\boldsymbol{m}=m_o\hat{\boldsymbol{z}}$, sitting at the origin of coordinates, is immersed in a time-independent magnetic field $\boldsymbol{H}(\boldsymbol{r})=H_x\hat{\boldsymbol{x}}+H_y\hat{\boldsymbol{y}}+H_z\hat{\boldsymbol{z}}$. According to Eq. (12a), the force on this dipole is given by

$$\boldsymbol{f}=(\boldsymbol{m}\cdot\boldsymbol{\nabla})\boldsymbol{H} = m_o(\partial/\partial z)(H_x\hat{\boldsymbol{x}}+H_y\hat{\boldsymbol{y}}+H_z\hat{\boldsymbol{z}}). \tag{15}$$

The Maxwell equations $\boldsymbol{\nabla}\times\boldsymbol{H}(\boldsymbol{r})=0$ and $\boldsymbol{\nabla}\cdot\boldsymbol{B}(\boldsymbol{r})=0$ yield $\partial H_x/\partial z=\partial H_z/\partial x$, $\partial H_y/\partial z=\partial H_z/\partial y$, and $\partial H_z/\partial z=-\partial H_x/\partial x-\partial H_y/\partial y$. Equation (15) may thus be rewritten

$$\boldsymbol{f}=(m_o\partial H_z/\partial x)\hat{\boldsymbol{x}}+(m_o\partial H_z/\partial y)\hat{\boldsymbol{y}}-m_o(\partial H_x/\partial x+\partial H_y/\partial y)\hat{\boldsymbol{z}}. \tag{16}$$

Now, the square current loop depicted in Fig. 2(a) experiences the Lorentz force of $H_z$ on its front and back legs, as well as that on its right and left legs, namely,

$$f_x=\mu_o[H_z(\tfrac{1}{2}\Delta,0,0)-H_z(-\tfrac{1}{2}\Delta,0,0)](I_o\Delta)=m_o\partial H_z/\partial x; \tag{17a}$$

$$f_y=\mu_o[H_z(0,\tfrac{1}{2}\Delta,0)-H_z(0,-\tfrac{1}{2}\Delta,0)](I_o\Delta)=m_o\partial H_z/\partial y. \tag{17b}$$

Similarly, the force of $H_x$ on the front and back legs plus the force of $H_y$ on the right and left legs is given by

$$f_z=\mu_o[H_x(-\tfrac{1}{2}\Delta,0,0)-H_x(\tfrac{1}{2}\Delta,0,0)](I_o\Delta)+\mu_o[H_y(0,-\tfrac{1}{2}\Delta,0)-H_y(0,\tfrac{1}{2}\Delta,0)](I_o\Delta)$$
$$=-m_o(\partial H_x/\partial x+\partial H_y/\partial y). \tag{17c}$$

The results of direct force calculation in Eq. (17) are seen to agree with the formal expression of force on the dipole given by Eq. (16). [A straightforward analysis reveals that the term $\boldsymbol{M}\times\boldsymbol{H}$ in the generalized torque expression of Eq. (12b) is similarly compatible with the current loop picture of the magnetic dipoles.] Consequently, the current loop interpretation may still be useful, at least in cases where the magnetic field is time-independent, even though, in general, the loop's exchanges of energy and momentum with the EM field do not hold up under close scrutiny.

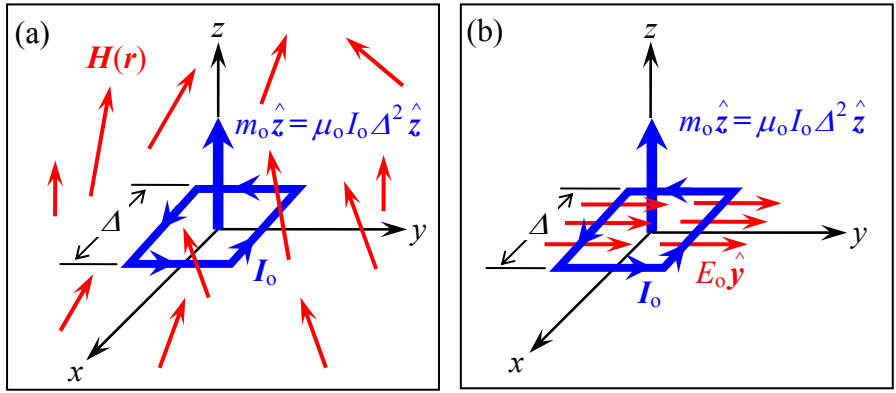

**Fig. 2**. (a) A small square loop of area $\Delta^2$ carries the constant current $I_o$ in the counterclockwise direction. The loop, being immersed in a time-independent magnetic field $\boldsymbol{H}(\boldsymbol{r})$, experiences a Lorentz force on each of its four legs by the action of the various components of the external field. (b) A time-independent, uniform electric field, $E_o\hat{\boldsymbol{y}}$, acts on a magnetic dipole whose Amperian current loop sits in the *xy*-plane at the origin of coordinates. Any change in the current $I_o$ (or in the loop's orientation) which would produce a change of the dipole moment *m*, gives rise to a force $\boldsymbol{f}=-(\partial\boldsymbol{m}/\partial t)\times\varepsilon_o\boldsymbol{E}$ on the dipole.

The term $-(\partial\boldsymbol{M}/\partial t)\times\varepsilon_o\boldsymbol{E}$ in the force density expression of Eq. (12a) is somewhat harder to explain. With reference to Fig. 2(b), consider a time-independent, uniform electric field $E_o\hat{\boldsymbol{y}}$ acting on a small, square current loop.

Suppose the current of the loop rises by $\delta I_o$ during the time interval $\delta t$. The front leg of the loop then absorbs energy from the $E$-field in the amount of $E_o(\delta I_o \Delta)\delta t$, while the rear leg returns the same amount of energy to the field during the same time interval. The mass associated with this energy is obtained by dividing the energy by $c^2$, and the corresponding momentum (propagating from front to back within the loop) is found by multiplying with the velocity $\Delta/\delta t$. We thus learn that, during the time interval $\delta t$, a net mechanical momentum, $\delta p = E_o \delta I_o \Delta^2/c^2$, has been shifted internally along the $-x$ direction. The backlash against this internal momentum transfer is observed externally as a force $f = \delta p/\delta t = E_o \delta I_o \Delta^2/(c^2 \delta t) = \varepsilon_o E_o(\mu_o \delta I_o \Delta^2)/\delta t$ acting on the current loop in the $+x$ direction. With $\delta \boldsymbol{m}$ being equal to $(\mu_o \delta I_o \Delta^2)\hat{z}$, the force is seen to be expressible as $\boldsymbol{f} = -(\partial \boldsymbol{m}/\partial t) \times \varepsilon_o \boldsymbol{E}$.

A similar argument can be made in the case of a rotating dipole, where the time-rate-of-change of the dipole moment $\boldsymbol{m}$ is not the result of a change in its circulating current, but rather a consequence of the turning of its loop in an external $E$-field.

Whereas the generalized force density expression of Eq. (12a) explicitly accounts for the internal momentum transfer of magnetic dipoles in the presence of an $E$-field, the standard Lorentz force expression, with its prescription of the $B$-field exerting a force on the bound electric current associated with magnetic dipoles, hides this momentum and gives rise to the discrepancies discussed in Sec. 6.

We emphasize that the above limited attempt at justifying the generalized Lorentz force expression of Eq. (12a), while perhaps useful for pedagogical purposes, is not at all needed for establishing the validity of the formula. Having established in previous publications [14-19], under the most general circumstances, the consistency of Eqs. (12a) and (12b) with the laws of linear and angular momentum conservation, this generalized version of the Lorentz law will now stand or fall depending on whether or not its predictions come to agree with experimental observations.

**10. Force exerted by electromagnetic fields on an electric dipole**. The foregoing analyses have been, in the main, concerned with magnetic dipoles. We would like to point out, however, that in their exchange of linear and angular momentum with the EM field, electric dipoles do not seem to behave in a conventional way either. In the fundamental expressions that define energy and force, namely, Eq. (5) and Eq. (12a), the bound-current density $\partial \boldsymbol{P}/\partial t$ of electric dipoles acts in the manner expected from an ordinary current density. However, the bound-charge density, $-\boldsymbol{\nabla} \cdot \boldsymbol{P}(\boldsymbol{r},t)$, does *not* play its expected role in the force equation. The generalized force density expression, Eq. (12a), contains the term $(\boldsymbol{P} \cdot \boldsymbol{\nabla})\boldsymbol{E}$, which is not the same as the expected term, $-(\boldsymbol{\nabla} \cdot \boldsymbol{P})\boldsymbol{E}$. Also, the term $\boldsymbol{P} \times \boldsymbol{E}$ in the generalized torque density formula, Eq. (12b), would have been absent had $\boldsymbol{P}(\boldsymbol{r},t)$ been fully reducible to bound-charge and bound-current densities.

We have shown elsewhere that, while the two expressions representing the conventional and generalized Lorentz force exerted on $\boldsymbol{P}(\boldsymbol{r},t)$, result in identical *total* force (and also *total* torque) experienced by solid objects, the force (and torque) *distributions* are quite different in the two formulations [32,33]. This could lead to measurable effects in soft bodies subjected to EM radiation. The fundamental quantum nature of the elementary electric dipole $\boldsymbol{p}$ is thus expected to assert itself in the macroscopic exchanges of linear and/or angular momentum between the EM field and flexible objects – objects that are responsive not just to the total force (or total torque) exerted upon them, but also to the distribution of force (or torque) throughout their volumes and across their surfaces.

**11. Concluding remarks**. Maxwell's macroscopic equations presuppose the existence, as well as the EM properties, of charge, current, polarization and magnetization, in the same way that Newton's laws of inertia and gravitation presuppose the existence and properties of matter. These equations cannot explain the internal structure of an electron, nor can they describe the constitution of an atom. But they say: "Assume there exist charge density distribution $\rho(\boldsymbol{r},t)$ and current density distribution $\boldsymbol{J}(\boldsymbol{r},t)$." They then proceed to describe the fields, $\boldsymbol{E}(\boldsymbol{r},t)$ and $\boldsymbol{B}(\boldsymbol{r},t)$, produced by these charges and currents. Similarly, the equations do not specify the nature of $\boldsymbol{P}$ and $\boldsymbol{M}$; these are simply given natural entities that give rise to, and also interact with, the EM fields. If the quantum nature of magnetic dipoles, for example, happens to endow them with certain rigidity and stability properties in the presence of external fields that would take them beyond the realm of classical Amperian loops, that fact will in no way violate or contradict the four equations of Maxwell.

The fact that Maxwell's equations have been successful in explaining the observed production of EM fields by $\rho_{\text{free}}$, $\boldsymbol{J}_{\text{free}}$, $\boldsymbol{P}$, and $\boldsymbol{M}$, means that these equations deal with "real-world" electric and magnetic dipoles, and not with some classical objects constructed, for instance, from an ordinary loop of wire that carries a certain amount of current. There is no need for Maxwell's equations to account, say, for the $g$-factor associated with the spin of an electron; the equations do not even pretend to explain the nature of magnetism. They simply postulate that $\boldsymbol{M}(\boldsymbol{r},t)$ exists in Nature. The equations then proceed to say that "$\mu_o^{-1} \boldsymbol{\nabla} \times \boldsymbol{M}$ appears in the same place in the equations as the density of an ordinary current, $J_{\text{free}}$." As a matter of fact, this last sentence is the source of *all* analogies between Amperian current loops and magnetic dipoles; there is no substance to the Amperian current loop above and beyond what is contained in this analogy, which, in turn, is exclusively based on Maxwell's equations themselves. Thus the notion of an Amperian current loop is of no consequence in the realm of Maxwell's equations, until and unless the analogy is pushed too far and brought into the realm of the auxiliary equations that describe the exchange of energy and momentum between fields and dipoles, which is what has given birth to the $\mu_o^{-1} E \times B$ form of the Poynting vector.

Recognizing that $\mu_o^{-1} \nabla \times M$ is a different kind of current, the kind that does *not* exchange energy with the *E*-field in the same way that $J_{\text{free}}$ and $\partial P/\partial t$ do, is the point of departure for the $E \times H$ form of the Poynting vector. In other words, when one uses the $S = E \times H$ formulation, one is automatically stating that $\mu_o^{-1} \nabla \times M$ is a special type of current, the type that neither picks up nor releases energy in the presence of an *E*-field. Conversely, adherence to the $S = \mu_o^{-1} E \times B$ form of the Poynting vector is tantamount to accepting $\mu_o^{-1} \nabla \times M$ as an ordinary current density, *J*, the kind that gives or takes energy in the presence of an *E*-field at the rate of $E \cdot J$.

Similar considerations apply to exchanges of momentum and angular momentum between the EM fields, on the one hand, and the electric and magnetic dipoles, on the other. If $\rho_{\text{e\_bound}} = -\nabla \cdot P$ and $J_{\text{e\_bound}} = \partial P/\partial t + \mu_o^{-1} \nabla \times M$ behaved as ordinary charge and current densities, then the force experienced by material media would be given by the conventional Lorentz force density, $F(r,t) = \rho_{\text{total}}(r,t) E(r,t) + J_{\text{total}}(r,t) \times B(r,t)$, while the torque density would be $T(r,t) = r \times F(r,t)$. This state of affairs, however, would give rise to hidden momentum and other absurdities, as discussed in the preceding sections. This paper has argued that avoiding the absurdities requires that the dipoles, in their interactions with EM fields, depart from the conventional Amperian model, which brings about the generalized force and torque density expressions given by Eqs. (12).

**Acknowledgements**. The author is grateful to Rodney Loudon, Armis Zakharian, Richard Ziolkowski, Ewan Wright, Henri Lezec, Pablo Saldanha, and George Schatz for helpful discussions. Special thanks are due to John Weiner for his careful review of the earlier drafts of the manuscript, and for his constructive criticism and substantive suggestions, which have resulted in significant improvements. The research reported here was conducted while the author was on sabbatical leave from the University of Arizona, serving as Visiting Chair Professor in the department of Physics at the National Taiwan University in Taipei; he thanks his host, Prof. Din Ping Tsai, for providing the opportunity and the friendly atmosphere in which to carry out this work.